\documentclass[preprint,aps,shortbibliography,natbib=true]{revtex4-2}
\usepackage[english]{babel}
\usepackage{physics}
\usepackage[letterpaper,top=2cm,bottom=2cm,left=3cm,right=3cm,marginparwidth=1.75cm]{geometry}
\usepackage{amsmath}
\usepackage{float}
\usepackage{graphicx}
\usepackage[dvipsnames]{xcolor}
\usepackage{bm}
\usepackage[colorlinks=true, allcolors=blue]{hyperref}
\usepackage[normalem]{ulem}

\begin{document}

\title{Spin Hall and Edelstein Effects in Novel Chiral Noncollinear Altermagnets}

%other possible title: Chiral Noncollinear Altermagnet: platform for multifunctional spintronics;
%Chiral Noncollinear Altermagnet: Mn$_3$IrSi

%\author{Mengli Hu}
%\affiliation{Nice places}
% \\$^{1}$Leibniz Institute for Solid State and Materials Research, IFW Dresden, Helmholtzstraße 20, 01069 Dresden, Germany\\
% $^2$Max Planck Institute for Chemical Physics of Solids, Dresden, Germany \\
% $^3$Donostia International Physics Center, Donostia-San Sebastian, Spain}

\author{Mengli Hu$^{1}$, Oleg Janson$^{1}$, Claudia Felser$^{2}$, Paul McClarty$^{3}$, Jeroen van den Brink$^{1,4,\sharp}$, Maia G. Vergniory$^{5,6,\sharp}$
}

\affiliation{
\\$^{1}$Leibniz Institute for Solid State and Materials Research, IFW Dresden, Helmholtzstraße 20, 01069 Dresden, Germany
\\$^{2}$Max Planck Institute for Chemical Physics of Solids, 01187 Dresden, Germany
\\$^{3}$Laboratoire Léon Brillouin, CEA, CNRS, Université Paris-Saclay, CEA-Saclay, 91191 Gif-sur-Yvette, France
\\$^{4}$Würzburg-Dresden Cluster of Excellence Ct.qmat, Technische Universitat Dresden, 01062, Dresden, Germany
\\$^{5}$ Département de Physique et Institut Quantique, Université de Sherbrooke, Sherbrooke,
J1K 2R1 Québec, Canada. 
\\$^{6}$Donostia International Physics Center, 20018 Donostia–San Sebastian, Spain
%\\$^{6}$Department of Physics, Stockholm University, AlbaNova University Center, 10691 Stockholm, Sweden
%\\$^{3}$Max Planck Institute for Chemical Physics of Solids, 01187 Dresden, Germany
%\\$^{9}$Donostia International Physics Center, 20018 Donostia–San Sebastian, Spain
%\\$^{*}$These authors contributed equally to the present work.
\\$^{\sharp}$Corresponding authors: j.van.den.brink@ifw-dresden.de, maia.vergniory@usherbrooke.ca
}

\begin{abstract}

Altermagnets are a newly discovered class of magnetic phases that combine the spin polarization behavior of ferromagnetic band structures with the vanishing net magnetization characteristic of antiferromagnets. Initially proposed for collinear magnets, the concept has since been extended to include certain non-collinear structures. A recent development in Landau theory for collinear altermagnets incorporates spin-space symmetries, providing a robust framework for identifying this class of materials. 
Here we expand on that theory to identify altermagnetic multipolar order parameters in non-collinear chiral materials.
We demonstrate that the interplay between non-collinear altermagnetism and chirality allows for spatially odd multipole components, leading to non-trivial spin textures on Fermi surfaces and unexpected transport phenomena, even in the absence of SOC. This makes such chiral altermagnets fundamentally different from the well-known SOC-driven Rashba-Edelstein and spin Hall effects used for 2D spintronics. 
Choosing the chiral topological magnetic material Mn$_3$IrSi as a case study, we apply toy models and first-principles calculations to predict experimental signatures, such as large spin-Hall and Edelstein effects, that have not been previously observed in altermagnets. These findings pave the way for a new realm of spintronics applications based on spin-transport properties of chiral altermagnets.

\end{abstract}
\maketitle

\section{Introduction}

Altermagnets (AMs) are compensated magnetic phases that share features of ferromagnets and antiferromagnets characterized by an alternation of magnetic moments synchronized with an alternation of local multipoles around magnetic atoms\cite{AM_1,AM_2}. 

In their original setting, altermagnets are collinearly ordered antiferromagnets that have zero net magnetization but yet a spin-splitting of electronics bands, also in the absence of relativistic spin-orbit coupling (SOC).
This leads to Fermi surfaces with non-trivial patterns of spin polarization in momentum space, for instance of $d$-wave type, that can sustain spin-currents \cite{AM_1,AM_2}, like ferromagnets while being robust to stray magnetic fields unlike their ferromagnetic counterparts. Such properties of altermagnets offer an intriguing opportunity to leverage the complementary advantages of ferro- and antiferromagnets in spintronics~\cite{AM_1,AM_2,AFMSpintronics1,AFMSpintronics2,p-wave_photocurrent,SOT_noncollinear}.
The recently developed Landau theory of collinear altermagnetism~\cite{AM_landau}, provides a yet more general, symmetry-based definition: an antiferromagnet is considered an AM when there is neither PT symmetry nor time reversal combined with a translation and when the N\'{e}el order parameter transforms nontrivially under the point group of the lattice %{\mgv Paul: which point symmetry? the one of the space/magnetic/Spin groups?} 
and leads to co-existing magnetic multipolar pseudo-primary order parameters. 
For more general magnetic orderings the presence of spin splitting requires, apart from the absence of PT symmetry, also the spin space PG of the spin translation group not to contain the dihedral group $D_n$\cite{ren2023enumeration,p-wave}.
The pseudo-primary AM order parameter is directly related to the spin splitting of the band structures\cite{AM_1,AM_2,AM_band2,AM_band1,CSVL_1,CSVL_2,multipole_2} and related key observables in, e.g., spin-transport.
%\cite{AM_landau}.
%
While it has been widely recognized that noncollinear spin ordering can be such that the moments are fully compensated, while still generating, e.g., an Anomalous Hall Effect (AHE) \cite{AM_noncollinear,dal2024antiferromagnetic,AHE_noncollinear,QHE_chiralmag,AM_AHE,AHE2010}, these observations in themselves do not provide an intrinsic connection to altermagnetism.

Here, we establish this connection by extending the scope of altermagnetic Landau theory to {\it noncollinear, chiral} systems. In addition to a symmetry-dictated vanishing of net magnetization we require for altermagnetism that the noncollinear N\'{e}el ordering induces a secondary, symmetry-induced multipolar order parameter. Unlike the basic definition, which relies solely on band spin-splitting—a generic feature of any parity-time broken (noncollinear) spin system—this definition connecting to multipolar order is rigorously grounded in symmetry, and, as a consequence, provides a direct connection to physical observables\cite{AM_landau,multipole_1}, as will be demonstrated in the following sections. 

To be specific we take the chiral magnet Mn$_3$IrSi, belonging to magnetic space group P2$_1$3~\cite{MIrSi_1,MIrSi_2,MIrSi_3}. Our symmetry analysis shows that its experimentally observed compensated noncollinear N\'{e}el order induces one monopolar and three quadrupolar altermagnetic secondary order parameters. These altermagnetic order parameters pass along a characteristic momentum-space
spin texture to the electronic structure, and the Fermi surface in particular. We determine the monopole and quadrupole symmetry components for Mn$_3$IrSi not only from full-scale first principles calculations, but also a  more general symmetry-appropriate magnetic model and establish their connection. 

While the physical consequence of the quadrupole spin texture is the spin polarization of charge currents, similar to the behavior observed in canonical collinear altermagnets, the monopole order parameter manifests itself through the bulk Edelstein and spin Hall effects. In this case, spin-momentum locking induces charge-spin conversion, resulting in an effective torque on the electrons. What sets these effects apart in such noncollinear chiral altermagnets is that they occur in the absence of SOC, making them fundamentally different from the SOC-driven linear and nonlinear Rashba-Edelstein~\cite{Edelstein_orbital,Edelstein_nonlinear,Edelstein_SC} and spin Hall effects and other non-trivial spin textures in chiral materials \cite{PhysRevB.106.245101,Krieger2024}. The key distinction is that the energy scale driving the altermagnetically induced transport mechanism is the spin-splitting of electronic bands, which is tied to the local magnetic exchange energy
\cite{AM_1,AM_2,AM_band1,AM_band2} and is orders of magnitude larger than the relativistic SOC. 
%
%The altermagnetism also stands out for its zero net moments compared with the Edelstein effect recently proposed in noncollinear magnets\cite{Edelstein_material1}. 
%
The altermagnetic symmetries that we consider here in addition constrain the net moment to zero, compared to generic noncollinear magnets \cite{Edelstein_material1} discussed recently for their spintronic properties.
Mn$_3$IrSi is an excellent candidate for observing these effects, as, despite the presence of iridium, our calculations show that SOC plays a negligible role in its electronic structure and altermagnetic spin texture.

This article is organized as follows: In Section \ref{secLT}, we elaborate on the Landau theory for non-collinear chiral altermagnets. Section \ref{secmirsi} is devoted to the electronic properties of Mn$_3$IrSi, where toy model and first-principles results are both shown. In Section \ref{sectrans}, we derive the unexpected transport properties resulting from the combination of chirality and noncollinear altermagnetism, which we determine quantitatively.
Finally, we conclude with a summary and conclusions in Section \ref{sumcon}.

\section{Landau theory of chiral noncollinear AM}\label{secLT}

In order to elaborate on the nature of altermagnetism in chiral, non-collinear compensated magnets, it is useful to first introduce altermagnetism in its original context, namely in collinear, compensated magnets with an inversion center in the crystal structure. With this foundation, we will then be in a position to discuss the new physics that emerges when the inversion center and collinearity are lost.

Altermagnetism is most conveniently defined in the limit of zero SOC. This is because there is a separation of energy scales in altermagnetic systems where the scale associated with spin splitting in the band structure is much larger than spin-orbit splittings.  With this in mind, we consider a lattice with some space group $G$. In the paramagnetic phase, the magnetism is symmetric under the elements of $G$ as well as rotations in spin space and time reversal. In formulating a Landau theory for such systems we introduce an order parameter $\boldsymbol{\Psi}$ corresponding to the collinear magnetic order that does not distinguish components in spin space. We may then define altermagnetism in such cases by requiring, first, that the magnetic sublattices are connected neither by an inversion nor a translation and, then, that the order parameter transform as a {\it non-trivial} one-dimensional irreducible representation (IR) $\Gamma$ of the point group of the lattice. This directly restricts to a set of crystal symmetries where the two magnetic sublattices are connected by a non-symmorphic rotation or mirror operation\cite{AM_landau}. The Landau theory is simply $F= c_2 \boldsymbol{\Psi}\cdot \boldsymbol{\Psi} + c_4 (\boldsymbol{\Psi}\cdot \boldsymbol{\Psi})^2$. With the condition on $\Gamma$, we may identify a spin symmetric, time odd, spatially anisotropic order parameter of the form, 
\begin{equation}
\boldsymbol{O}_\Gamma = \int d^3\mathbf{r} [ r_{\mu_1} \ldots r_{\mu_p}] \boldsymbol{s}(\mathbf{r}) 
\end{equation}
that transforms like $\Gamma$ and where $\boldsymbol{s}(\mathbf{r})$ is the local magnetization density and $[\ldots]$ denotes a symmetrization operation. As this transforms like $\Gamma$ it must enter the Landau theory through the additional term $\lambda \boldsymbol{\Psi}\cdot \boldsymbol{O}$. Therefore $\boldsymbol{O}$ is a secondary (or pseudo-primary) order parameter. Interestingly this multipolar order parameter is tied directly to the anisotropy in the spin structure of the Fermi surface. For example, in rutile crystals with chemical formula MX$_2$ where M is magnetic with a sublattice N\'{e}el order parameter, and point group $D_{4h}$, the relevant multipolar order parameter is $\int d^3\mathbf{r} xy \boldsymbol{s}(\mathbf{r})$ implying that the band  structure exhibits a $d$-wave spin splitting $-$ a rotation from $k_x$ to $k_y$ reverses the spin. This result is borne out by {\it ab initio} calculations of rutile magnets\cite{RuO2_Hall,AM_2,AM_band1,bhowal2024}. 

The Landau theory further exemplifies an important aspect of collinear altermagnets namely that the order parameter breaks down a paramagnetic spin symmetry group to a {\it collinear spin group}. Spin groups enlarge the set of magnetic symmetry groups by including elements that do not transform spin and space identically \cite{brinkman1966,litvin1974,corticelli2022,schiff2023,Liu,yang2023symmetry,xiao2023spin,ren2023enumeration,jiang2023enumeration}. For example, the rutile magnetic order includes an element with $C_2$ in spin space, reversing the moment, together with a composition of a $C_4$ and a translation in real space that swaps the magnetic sublattices. Collinear spin groups also include elements only acting on spin including global rotations about the moment direction and $C_2 T$ where $T$ is time reversal. 

These concepts may be naturally generalized to non-collinear magnetic structures \cite{AM_landau}. Again, one may define altermagnetism to correspond to magnetic order parameters at zero SOC that transform non-trivially under the point group of the lattice but now no longer restricting to 1D IRs. We point out that an important difference compared to the collinear case is that bands no longer carry a global spin quantum number. Instead, there is a well-defined spin at each momentum forming a spin texture across, for example, a Fermi surface. But the nature of the momentum-space spin texture can be characterized by the same kind of multipolar order parameters we described in the collinear case.

We now focus on altermagnets in magnetic chiral crystals. It has been pointed out that collinear chiral crystals retain the $C_2 T$ pure spin symmetry of their achiral counterparts. This operation takes $\mathbf{k}\rightarrow - \mathbf{k}$ while preserving the spin. This means that there is an effective inversion symmetry. Noncollinear altermagnets stand out because they break this $C_2 T$ and the chirality can therefore be manifest in momentum space. Specifically, this means that the real space part of the multipolar order parameter can be odd. We shall see how this feature is reflected in the spin texture in momentum space. We shall also see that this leads to experimental spin-transport signatures that stand apart from altermagnets studied to date. 

To illustrate altermagnetism in chiral noncollinear systems, we consider Mn$_3$IrSi, which is a particularly interesting magnetic system due to its multifold topological semimetal properties\cite{MIrSi_multifold,MultipleFermion,APL_MGV,magnetic_topology}. Moreover, it belongs to the family of $\beta$-Mn type alloys: Mn$_3TX$ ($T$ = Co, Rh, and Ir; $X$ = Si and Ge), all of which share the same crystal and magnetic structure. Various features of these materials have been reported in the literature, including short-range magnetic ordering\cite{Mn3RhSi_shortrange,Mn3CoSi}, an incommensurate magnetic phase\cite{MIrSi_MRhGe} at high temperature, and a doping-induced magnetic phase transformation\cite{MnCoGe}. 

\begin{figure*}[h]
  \includegraphics[width=0.9\textwidth]{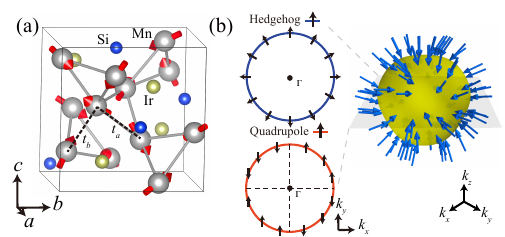}
  \caption{\label{cystal_monoqua} Crystal structure of Mn$_3$IrSi and the momentum space spin texture from Landau theory a model calculation.
(a) The Mn$_3$IrSi unit cell contains 12 Mn atoms, with local magnetic moments indicated. Both the crystal and magnetic structures follow chiral symmetry, and the magnetic moments are noncollinear. The bonding between nearest and second-nearest Mn atoms is labeled by dashed black lines.
(b) The hedgehog and quadrupole spin textures predicted in the chiral noncollinear altermagnet. They are inversion odd: $\textbf{s(k)} = -\textbf{s(-k)}$ and inversion even: $\textbf{s(k)} = \textbf{s(-k)}$, respectively. The three-dimensional Fermi surface is taken from the minimal-band toy model of Mn$_3$IrSi at $E = E_f + 0.08$ eV. 
The schematic plot shown in the left panel is taken at the $k_x-k_y$ plane in grey.
  }
  \end{figure*}

As shown in Fig.~\ref{cystal_monoqua}a, Mn$_3$IrSi has 12 Mn atoms with noncollinear local magnetic moments belonging to the same unit cell. The local magnetic direction is denoted by red arrows. The space group of Mn$_3$IrSi crystal is P2$_1$3 (No. 198) which belongs to the Sohncke space groups, with the magnetic atoms occupying the Wyckoff position $12b$.  It can be described by both the magnetic space group (MSG)  P2$_1$3  and spin space group (SSG) and P$^{2_{100}}$2$_1^{3^1_{111}}3$, which, as we will see, are isomorphic. Due to the noncollinear and noncoplanar properties of its magnetic structure, the spin-only group is trivial, and the point group symmetry operations are identical in both real space and the spin basis. In other words, because of the noncollinear altermagnetic (AM) nature, the spin space group (AM group) and the magnetic space group of Mn$_3$IrSi are isomorphic. 

Thus, although SOC is strictly zero in the Hamiltonian, the non-collinear magnetic ordering causes the magnetic symmetries to be equivalent to those expected in a finite spin-orbit coupled system. Nevertheless in contrast to the original collinear altermagnets, inversion symmetry is broken. However, Mn$_3$IrSi shares a key feature with collinear altermagnets: the absence of spin degeneracy, leading to the expectation of a spin-split band structure. 

In order to understand this spin splitting from a symmetry perspective, we begin with the 12-dimensional representation based on the 12b sublattice basis. This representation can be decomposed under the tetrahedral point group $T$ as: $A\oplus E \oplus 3T$. As the experimentally observed magnetic order carries zero total magnetization
($\bm{M} = 0$), which is also the magnetic ground state found in first-principles calculations, we can exclude a magnetic order parameter associated with the irreducible representation (IR) $A$, which corresponds to the total magnetization: $\bm{M} = \sum_{i\in prim}\sum_{a=1}^{12}\bm{S}_{ia}$.
For the IR $E$, the magnetic structure of Mn$_3$IrSi is orthogonal to its components, and $\bm{\Phi}^{\alpha}=0$, where $\bm{\Phi}^{\alpha} = \sum^{12}_{a=1} \phi^{\alpha}_{E,a}\bm{S}_a$, and $\forall \alpha$, $\bm{\Phi}^{\alpha}=0$ due to the orthogonal properties. Thus, the only remaining order parameter is related to the IR $T$ (see also Ref.~\cite{xiao2023spin}). 
After projecting the 12-dimensional vector onto the IR $T$, the basis of $T$ reads: $\phi^1_T = (x,-x,-x,x,y,y,-y,-y,z,-z,z,-z); \phi^2_T = (z,-z,z,-z,x,-x,-x,x,y,y,-y,-y);\phi^3_T = (y,y,-y,-y,z,-z,z,-z,x,-x,-x,x)$. Combined with the magnetic structure: $\bm{\Phi}^{\alpha} = \sum^{12}_{a=1}\phi^{\alpha}_{T,a}\bm{S}_a$. Based on the  order parameter of $T$, we can write down the associated Landau theory:
\begin{equation}
F = c_2 \bm{\Phi}^{\alpha}\cdot \bm{\Phi}_{\alpha} + c_4 ( \bm{\Phi}^{\alpha}\cdot \bm{\Phi}_{\alpha})^2.
\end{equation}
As the total magnetization transforms as $A$, there is no direct coupling between the magnetic order parameter and $\bm{M}$. But other variables related to $T$  can couple to $\bm{\Phi}^{\alpha}$, in particular, the spatial dipole and quadrupole terms defined as:
\begin{equation}
\begin{aligned}
\bm{L}_{1,\alpha} &= \int d^3r r_{\alpha} \bm{m}(\bm{r}) \\
\bm{L}_{2,\alpha} &= \int d^3r Q_{\alpha} \bm{m}(\bm{r}) \ \ Q_{\alpha} = (r_xr_y,r_xr_z,r_yr_z).
\end{aligned}
\end{equation}

On the basis of the first term in momentum space, a hedgehog spin texture is expected, where the local spin at $\bm{k}$ ($\bm{s}(\bm{k})$) reverses sign under time-reversal symmetry: $\bm{s}(\bm{k}) = -\bm{s}(-\bm{k})$. For the quadrupole spatial term, this couples to a quadrupolar spin texture, which transforms differently under time-reversal symmetry: $\bm{s}(\bm{k}) = \bm{s}(-\bm{k})$. 
Fig.~\ref{cystal_monoqua}b illustrates both the hedgehog and quadrupole-like two-dimensional spin textures, with the quadrupolar component showing zero spin at the crossing points along the $k_x$ and $k_y$ axes. The predicted spin texture, which belongs to the IR $T$, aligns with previous studies on spin space groups\cite{xiao2023spin}. 

In the original Landau theory of altermagnetism (AM)\cite{AM_landau}, an antiferromagnet is considered an altermagnet because its N\'{e}el order parameter transforms non-trivially under point group symmetries, resulting in co-existing magnetic multipolar pseudo-primary order parameters. A key characteristic of chiral crystals is the absence of inversion and mirror symmetries, meaning all improper rotational symmetries are absent. Consequently, spatial and axial vectors transform identically and can belong to the same IRs. For the chiral noncollinear altermagnet Mn$_3$IrSi, the non-trivial magnetic order parameter $\bm{\Phi}^{\alpha}$ is established, and as a consequence, finite magnetic multipolar order parameters $\bm{L}_{1,\alpha}$ and $\bm{L}_{2,\alpha}$, along with their corresponding spin textures in momentum space are expected on the basis of symmetry.

Finally, we consider the effect of SOC on symmetry under the assumption that altermagnetism persists when the spin-orbit splittings are smaller than the spin-splitting in the absence of SOC. When SOC is introduced, the 12 sublattices, along with the 3 local spin components for each, must be considered. This results in a 36-dimensional representation. The spin-symmetric basis in the $T$ IR now breaks into three copies of the $A$ IR, which reads:
\begin{equation}
\begin{aligned}
    \phi_{A1} &= -S^z_1-S^z_2+S^z_3+S^z_4-S^x_5-S^x_6+S^x_7+S^x_8-S^y_9-S^y_{10}+S^y_{11}+S^y_{12} \\
    \phi_{A2} &= -S^z_1+S^z_2+S^z_3-S^z_4-S^x_5+S^x_6+S^x_7-S^x_8-S^y_9+S^y_{10}+S^y_{11}-S^y_{12} \\
    \phi_{A3} &= -S^z_1+S^z_2-S^z_3+S^z_4-S^x_5+S^x_6-S^x_7+S^x_8-S^y_9+S^y_{10}-S^y_{11}+S^y_{12},
\end{aligned}
\end{equation}
whereas the magnetization transforms as $T$. 
Because the antiferromagnetic order parameter and magnetization transform differently (as A and T respectively) there can be no linear coupling between the two in the Landau theory. So to linear order there is no weak ferromagnetism (no small induced moment) and, as the anomalous Hall effect transforms like the magnetization, this too is not switched on to linear order. 

\section{Electronic properties of Mn$_3$IrSi}\label{secmirsi}

In the previous section, we demonstrated on symmetry grounds that the non-collinear magnetically ordered phase of Mn$_3$IrSi must exhibit a spin texture on its Fermi surface, with monopolar and quadrupolar spatial components, even in the absence of SOC. In this section, we first develop a simple Kondo-lattice model that with only few parameters accounts for the non-collinearly ordered magnetic moments. From it the Fermi surface spin textures predicted by Landau theory are identified. Introducing SOC as a perturbation does not significantly affect the spin texture, providing a robust minimal model for the chiral non-collinear altermagnet. Next, we present the first-principles results for Mn$_3$IrSi, showing that the spin texture aligns with the predictions, though it is more complex than our simplified toy model. Additionally, we demonstrate that the band structures with and without SOC are quite similar, indicating that SOC remains a small perturbation, despite the presence of the heavy element iridium. In the following section, both the toy model and the projected Wannier functions of Mn$_3$IrSi will be used to predict novel, robust transport phenomena that do not rely on the presence of SOC.

\begin{figure*}[h!]
  \includegraphics[width=1.0\textwidth]{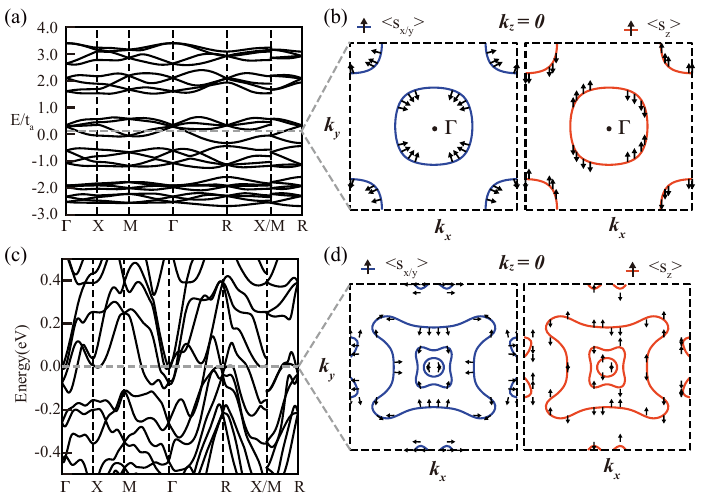}
  \caption{\label{band} Band structure and spin texture of Mn$_3$IrSi from Kondo-lattice model and from first-principles. (a) and (c) are the band structure along high symmetry lines from toy model and first-principles calculation, respectively. (b) and (d) are the spin texture at $E = E_f +0.08$ eV and Fermi level ($E_f$) from first-principles and toy model calculation, respectively. The left panel shows the spin components of $<\sigma_{x/y}>$ and the right panel presents the spin components of $<\sigma_z>$. All spins are normalized for clearer visualization.
  }
  \end{figure*}  

To capture the magnetic symmetry in Mn$_3$IrSi and other compounds with the same crystal structure, we propose a minimal toy model of tight-binding electrons hopping between the Mn sites on the $\beta$-Mn lattice, where magnetism is introduced via a Kondo-like coupling to fixed classical magnetic moments. These moments are arranged according to the experimentally observed non collinear
magnetic structure of Mn$_3$IrSi. The model is isotropic in spin space. To account for the effects of SOC, an additional symmetry allowed spin-dependent hopping term is included \cite{AM_landau,minimalmodels2024}. The Hamiltonian of the toy model is given by:
\begin{equation}
    \hat{H} = \sum_{\alpha,<<i,j>>}t_{ij}a^{\dagger}_{i,\alpha}a_{j,,\alpha} + \sum_i \bm{m}_i \cdot a^{\dagger}_{i,\alpha}\bm{\sigma}_{\alpha \beta} a_{i,\beta} + \sum_{<i,j>} \bm{s}_{ij} \cdot a^{\dagger}_{i,\alpha}\bm{\sigma}_{\alpha \beta} a_{j,\beta} + h.c.,
\end{equation}
where $t_{ij}$ represents the hopping parameters up to the second-nearest neighbors, $\bm{m_i}$ denotes the local magnetic moment strength, and $\bm{s_{ij}}$ is the off-site SOC strength. The Mn sublattice consists of 12 atoms, and a basis for the states is chosen to be $\ket{\alpha,\sigma}$, where $\alpha$ and $\sigma$ represent the sublattice and spin degrees of freedom, respectively. We cut off the hopping at the second-nearest neighbors ($t_i, i=a,b$) and limit the SOC to the nearest neighbors ($\textbf{s}$). 
Fig.~\ref{band}a shows the calculated band structure of the toy model along high symmetric lines with parameters $t_b/t_a=0.5$, $\textbf{s}/t_a = (0.05,0.03,0.02)$. Without loss of generality, the local magnetic moments are chosen as $m_1/t_a=0.3*(1.640, 2.774, -2.231)$. Fig.~\ref{band}b illustrates momentum-space spin texture arising in this chiral noncollinear altermagnetic model.
Only two bands are involved around Fermi level and their spin texture exhibits both the hedgehog winding spin texture around $\Gamma$ and $M$ points in $s_{x/y}$ and the quadrupole-like spin texture in the $s_z$ component. 

Having established the predicted momentum space spin texture is a generic feature of in the toy model, we proceed to calculate the detailed electronic structure of Mn$_3$IrSi.
As shown in Fig.~\ref{band}c, the band structure along high-symmetry lines agrees well with previous reports \cite{MIrSi_multifold}. The multifold degeneracies at the $R$ point are robust against the introduction of SOC. In the supplementary material, we compare the band structures, shown in figure S5, with and without SOC, concluding that there is minimal difference between them.

Fig.~\ref{band}d shows the spin texture without SOC revealing a hedgehog and quadrupole component in $\sigma_x$ and $\sigma_y$, and $\sigma_z$, respectively, in the $k_z=0$ plane.
The spin textures at different $k_z$ values are presented in figure S1, all showing a hedgehog-like component. In addition, the spin texture in $\sigma_z$ exhibits a quadrupole-like distribution, which is even under the time-reversal operation: $s_z(\bm{k}) = s_z(-\bm{k})$. This quadrupole-like spin texture is also a consequence of the primary order parameter in collinear altermagnetism (AM)\cite{AM_landau}.

One may also understand the presence of the hedgehog and quadrupole spin textures in momentum space from the basic symmetries of chiral noncollinear AMs. Such AMs are only allowed to have rotational symmetries that relate a group of states as: $\epsilon(\bm{k}_i) = \epsilon_0, \bm{s}(\bm{k}_i) = R_i(\theta_i)\bm{s}(\bm{k}_0),  \bm{k}_i\in R_i(\theta_i)\bm{k}_0$, where $R(\theta)$ is a rotation operator. From both symmetry analysis and DFT calculations of Mn$_3$IrSi, in the $k_z=0$ plane, $\bm{s}(\bm{k}_i) = R_i(\pi/2)\bm{s}(\bm{k}_0),i = x,y,z$, where $s_x, s_y$, and $s_z$ transform as dipole, and quadrupole, respectively.This spin-momentum locking relationship leads to both hedgehog-like and quadrupole-like spin textures. Given the good agreement between the toy model and the realistic material calculations, we conclude that our model may be easily extended for more general applications and analyze hedgehog-like and quadrupole-like spin textures in more classes of related chiral noncollinear altermagnets.
  
\section{Spin Hall and Edelstein effects}\label{sectrans}

We now show that the chiral noncollinear altermagnet discussed above carries both a spin Hall effect and an Edelstein effect as a consequence of the spin texture on the Fermi surface {\it even in the absence of SOC}. Both phenomena arise in the presence of an externally applied electric field: the appearance of a transverse spin current signalling the spin Hall effect while the Edelstein effect corresponds to a net magnetization. More precisely, the spin current $\mathcal{J}_j^i$ for spin component $i$ and current direction $j$ may depend on the electric field $E_k$ through $\mathcal{J}_{j}^{i} = \sum_{k} \sigma_{jk}^{i}E_{k}$ and the spin Hall effect corresponds to terms off-diagonal in $jk$. The Edelstein effect is a change in magnetization $\delta m_i = \chi^s_{ij}E_{j}$. Both phenomena are well-established in various systems in the presence of SOC \cite{SpinHall1,Edelstein_ori}. 

For the crystal symmetry of Mn$_3$IrSi, the spin Hall tensor $\sigma_{jk}^i$ has two independent, time-reversal even, non-vanishing components: $\sigma_{xy}^z$ and $\sigma_{xz}^y$ each transforming as the $A$ IR of the group. These therefore directly couple to the squared order parameter in the presence of SOC as the order parameter also transforms like $A$. In the spin-orbit free case, we saw that the order parameter transforms instead like $T$. As $T\otimes T$ contains $A$, there is a component that produces a spin Hall effect. 

A prerequisite for a non-vanishing Edelstein effect is that the crystal should not have inversion symmetry. For the crystal structure of Mn$_3$IrSi, the Edelstein tensor has a single diagonal component $\mathbf{m} = \chi^S \mathbf{E}$ that transforms like $A$ and is time reversal odd. This couples linearly to the order parameter in the presence of SOC. In the altermagnetic case of zero SOC there is no linear coupling. Instead, the Edelstein effect is {\it cubic} in the order parameter as $T\otimes T\otimes T = 2A \oplus \ldots$. 

\begin{figure*}[htpb!]
  \includegraphics[width=1.0\textwidth]{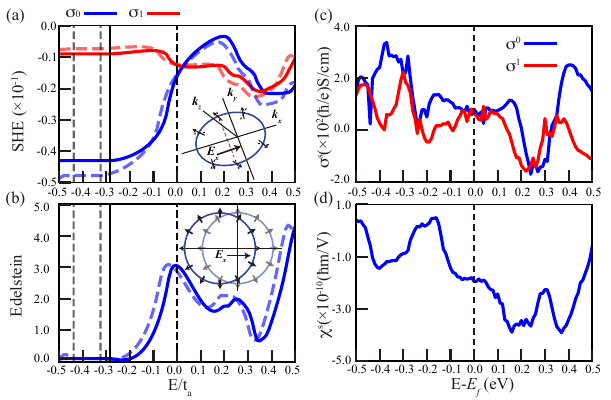}
  \caption{\label{SHE} Spin Hall and Edelstein effect conductivity tensor element ($\sigma_{1/0} = (\sigma^z_{xy} \pm \sigma^y_{xz})/2$) versus energy around Fermi level. (a) and (c) are spin Hall effect results from the model and projected Wannier functions of Mn$_3$IrSi, respectively. (b) and (d) are the Edelstein effect results ($\chi^s$) with the inverse scattering time $\Gamma^2= 10^{-4}$ ($t_a$) and (eV), respectively. The light dashed line in (a) and (b) are from the toy model without SOC.}
  \end{figure*}

Having seen that a spin Hall effect and Edelstein effect are allowed on symmetry grounds, we now show that they arise in Mn$_3$IrSi directly from a microscopic calculation. For spin current operator $\hat{A}^i_{j} = \mathcal{J}^i_j= \frac{1}{2}\{\hat{s}_i,\hat{v}_j \}$ we compute the linear spin Hall response \cite{Zelezny2017,Freimuth2014} with a constant inverse scattering time $\Gamma$. Two components contribute to the observable $\delta \hat{A}^i_j = (\chi_{i,jk} ^{I} + \chi_{i,jk} ^{II})E_k $, where:
\begin{equation}\label{kubo}
\begin{aligned}
\chi ^{I}_{i,jk} &= -\frac{e\hbar}{\pi VN}\sum_{\bm{k},m,n} \frac{\Gamma^2 Re (\mel{n\bm{k}}{\hat{A}^i_{j}}{m\bm{k}}\mel{m\bm{k}}{\hat{v}_k}{n\bm{k}})}{[(E_f - \epsilon_{n\bm{k}})^2 + \Gamma^2][(E_f - \epsilon_{m\bm{k}})^2 + \Gamma^2]} , \\
\chi ^{II}_{i,jk} & = -\frac{2\hbar e}{VN} \sum_{\bm{k},n\neq m}^{\substack{n=occ \\ m=unocc}} \frac{ Im (\mel{n\bm{k}}{\hat{A}^i_{j}}{m\bm{k}}\mel{m\bm{k}}{\hat{v}_k }{n\bm{k}})}{(\epsilon_{n\bm{k}} - \epsilon_{m\bm{k}})^2}.
\end{aligned}
\end{equation}
Here $e$ is the elementary charge, $\bm{k}$ is the Bloch wave vector, $n, m$ are the band indices, $\epsilon_{n,\bm{k}}$ is the eigenvalue, $E_f$ is the Fermi energy, $\hat{\bm{v}}$ is the velocity operator, 
% $\bm{E}$ is the direction of the electric field, 
$N$ is the total number of Bloch wave, and $V$ is the volume of unit cell. 
In the expression for $\chi^{II}$, the ranges of $m$ and $n$ refers to all the occupied bands and unoccupied bands, respectively, which is analogous to the calculation of Berry curvature. The intrinsic spin Hall effect (SHE) is defined to be the antisymmetric part from $\chi^{II}$\cite{SpinHall1} and is shown in Fig.~\ref{SHE}. The Edelstein effect results are evaluated from $\chi^{I}$ with $\hat{A}_i = \hat{s}_i$. 

Fig.~\ref{SHE} shows the calculated SHE (upper panels) and Edelstein effect (lower panels) for both the toy model (left-hand-side) and the Mn$_3$IrSi first-principles calculations (right-hand-side). The calculations are presented at zero temperature as a function of energy relative to the Fermi level. While the magnitude and even sign of the responses are parameter dependent, the calculations clearly reveal the presence of both a spin Hall current and electric field induced magnetization except at fine-tuned energies.

The microscopic calculations supply useful intuition connecting the spin texture at the Fermi level to the observables. The physical mechanism is illustrated in the insets of Fig.~\ref{SHE}(a) and (b), where the electric field displaces the Fermi surface in reciprocal space. The hedgehog spin texture then ensures that spins at the Fermi surface no longer compensate thus leading directly to an induced magnetization. A similar displacement causes electrons to experience an effective torque and tilt towards the $\sigma_z$ direction. A net spin current perpendicular to electric field ($E_x$) appears due to the opposite sign of the torque between momentum $k_y>0$ and $k_y<0$ as indicated in the inset to panel (a).

Quantitatively, compared to other predicted spin Hall materials, Mn$_3$IrSi has a relatively large SHE of $\sim 10^2 (\hbar /e)S/cm$~\cite{SHE_high}. This is an order of magnitude larger than SOC induced intrinsic values calculated for, e.g., nonmagnetic GeAs, AlAs or Ge~\cite{SHE_cal} and comparable to the predicted SHE in collinear  antiferromagnets~\cite{SHE_mag_cal,SHE_mag_cal2}. The value of the Edelstein effect that is caused by the chiral, noncollinear altermagnetism in Mn$_3$IrSi is of the same order of magnitude as that of certain noncoplanar magnets~\cite{Edelstein_material1}.

\section{Summary and conclusions}\label{sumcon}

Using group theory and Landau theory, we have predicted the existence of non-collinear chiral altermagnets and their distinctive electronic and transport properties. Compared to collinear altermagnets, non-collinear systems exhibit a more intricate momentum-space spin texture, extending the classification scheme  for collinear systems. On theoretical grounds, we have established the presence of a unique spin-momentum locking mechanism that arises in the absence of SOC, as a direct consequence of the chiral altermagnetism in Mn$_3$IrSi. This generalizes altermagnetism from its original context in collinear magnets $-$ where entire electronic bands can be labeled by a common spin quantum number $-$ to systems where the bands possess a local spin degree of freedom that is globally constrained to form a momentum-space texture governed by symmetry. This spin texture has direct physical implications. 
Due to the stricter symmetry constraints in non-chiral altermagnets (AM), spatially odd multipole components are not allowed in either collinear or non-collinear magnetic structures \cite{AM_landau}. Thus, chirality emerges as one of the necessary conditions for the hedgehog spin texture in AM. As a consequence, as exemplified in Mn$_3$IrSi, large spin Hall ($\sim 10^2 (\hbar /e)S/\text{cm}$) and Edelstein ($\sim -2 \times 10^{-10} \hbar ,\text{m/V}$) effects co-exist and have been calculated, showing very weak SOC dependence. 
We foresee multifunctional applications of chiral non-collinear AMs in spintronics and suggest additional candidates from the Mn$_3$IrSi family: Mn$_3$IrGe\cite{MnIrGe}, Mn$_3$CoGe\cite{MnCoGe}, Mn$_3$RhGe\cite{MIrSi_MRhGe}, Mn$_3$IrGe\cite{MnIrGe}; and other chiral noncollinear AMs: ScMnO$_3$\cite{ScMnO3}, BaCuTe$_2$O$_6$\cite{BaCuTe2⁢O6_1,BaCuTe2⁢O6_2}, YMnO$_3$\cite{YMnO3}, Ho$_2$Ge$_2$O$_7$\cite{Ho2Ge2O7}, Er$_2$Ge$_2$O$_7$\cite{Er2Ge2O7}.

\section{Methods}

The first-principles calculations were performed using VASP \cite{vasp}, employing the projector augmented wave (PAW) method \cite{paw}. The Brillouin zone was sampled on a $7\times7\times7$ $k$-point grid centered on the Gamma point. The energy cutoff for the plane wave basis was set to 550 eV. The Hubbard term was introduced with a value of 3.0 eV for the $d$ orbitals of the Mn atoms within the DFT+U framework to account for electron-electron correlations. The Wannier-based Hamiltonian was symmetrized based on the maximally localized Wannier functions generated by the WANNIER90 interface \cite{wannier}. The projectors were the $d$ orbitals of Mn, with the fitted region spanning from -2 to 2 eV. The magnetic moments within the self-consistent noncollinear ground state, with and without SOC, are respectively: $\abs{\bm{m}} = 3.91\pm 10^{-2} \mu_B$ and $\bm{m} = (1.640,2.774,-2.231)\pm 10^{-3} \mu_B$ on the Mn atom located at $(0.1195, 0.2031, 0.4573)$ in fractional coordinates of the unit cell.

\section{Acknowledgements} 
M.G.V. acknowledges fruitful discussions to Fernando de Juan. 
M.G.V. thanks support to the Deutsche Forschungsgemeinschaft (DFG, German Research Foundation) GA 3314/1-1 – FOR 5249 (QUAST), the Spanish Ministerio de Ciencia e Innovación (PID2022-142008NB-I00),  the Canada Excellence Research Chairs Program for Topological Quantum Matter and funding from the IKUR Strategy under the collaboration agreement between Ikerbasque Foundation and DIPC on behalf of the Department of Education of the Basque Government.
M.H. thanks Ulrike Nitzsche for technical assistance. 
M.H. thanks the support from the Alexander von Humboldt Foundation.
We acknowledge financial support by the Deutsche Forschungsgemeinschaft (DFG, German Research Foundation), through SFB 1143 (Project ID 247310070), project A05, Project No. 465000489, and the W\"urzburg-Dresden Cluster of Excellence on Complexity and Topology in Quantum Matter, ct.qmat (EXC 2147, Project ID 390858490).

% \bibliography{noncol_AM}
%apsrev4-2.bst 2019-01-14 (MD) hand-edited version of apsrev4-1.bst
%Control: key (0)
%Control: author (8) initials jnrlst
%Control: editor formatted (1) identically to author
%Control: production of article title (0) allowed
%Control: page (0) single
%Control: year (1) truncated
%Control: production of eprint (0) enabled
%

\end{document}